\documentclass[3p,11pt,onecolumn,sort&compress]{elsarticle}

\usepackage{amsfonts}
\usepackage{amsmath}
\usepackage{graphicx}
\usepackage{dcolumn}
\usepackage{bm}
\usepackage{hyperref}
\usepackage{cleveref}
\usepackage{subcaption}
\usepackage{placeins}
\usepackage{todonotes}
\usepackage{xspace}

\usepackage{array}   
\newcolumntype{L}{>{$}l<{$}} 
\newcolumntype{R}{>{$}r<{$}}
\newcolumntype{C}{>{$}c<{$}}
\usepackage{color}
\usepackage{braket}
\usepackage{units}
\usepackage{slashed}
\usepackage{soul}

\newcommand{\abs}[1]{\ensuremath{|#1|}}

\newcommand{\propagator}[1]{\mathcal{P}_{#1}}

\newcommand{\pomeron}{\mathbb{P}}
\newcommand{\units}[1]{\ensuremath{\text{ #1}}\xspace}

\newcommand{\elab}{{\ensuremath {E_\text{lab}}}}

\newcommand{\dsigdt}{\ensuremath{\frac{\text{d}\sigma}{\text{d}t}}}
\newcommand{\dsigdtperppara}{\ensuremath{\frac{\text{d}\sigma_{\perp/\parallel}}{\text{d}t}}}
\newcommand{\dsig}{\text{d}\sigma}
\renewcommand\Re{\operatorname{Re}}

\newcommand{\betahat}{\ensuremath{\hat\beta}}
\newcommand{\txdsigdt}{\ensuremath{\text{d}\sigma/\text{d}t}}
\newcommand{\txdsigdtperppara}{\ensuremath{\text{d}\sigma_{\perp/\parallel}/\text{d}t}}

\journal{Physics Letters  B}
\usepackage{hyperref}

\begin{document}

\begin{frontmatter}

\title{Features of $\pi \Delta$ Photoproduction at High Energies}

\author[ghent,IU,CEEM,jlab,GWU]{J.~Nys \corref{cor1}}
\ead{jannes.nys@ugent.be}
\author[jlab]{V.~Mathieu}
\author[UNAM]{C.~Fern\'andez-Ram\'irez}
\author[IU,CEEM]{A.~Jackura}
\author[Bonn]{M.~Mikhasenko}
\author[jlab]{A.~Pilloni}
\author[IU,CEEM]{N.~Sherrill}
\author[ghent]{J.~Ryckebusch}
\author[IU,CEEM,jlab]{A.~P.~Szczepaniak}
\author[IUinfo]{G.~Fox}

\address[ghent]{Department of Physics and Astronomy, Ghent University, B-9000 Ghent, Belgium}
\address[IU]{Physics Department, Indiana University, Bloomington, IN 47405, USA}
\address[CEEM]{Center for Exploration of Energy and Matter, Indiana University, Bloomington, IN 47403, USA}
\address[jlab]{Theory Center, Thomas Jefferson National Accelerator Facility, Newport News, VA 23606, USA}
\address[GWU]{The George Washington University, Washington, DC 20052, USA}
\address[UNAM]{Instituto de Ciencias Nucleares, Universidad Nacional Aut\'onoma de M\'exico, Ciudad de M\'exico 04510, Mexico}
\address[Bonn]{{Universit\"at Bonn, Helmholtz-Institut f\"ur Strahlen- und Kernphysik, 53115 Bonn, Germany}}
\address[IUinfo]{School of Informatics, Computing, and Engineering, Indiana University, Bloomington, IN 47405, USA}

\author{\\[.4cm] (Joint Physics Analysis Center)}

\cortext[cor1]{Corresponding author}

\begin{abstract}
Hybrid/exotic meson spectroscopy searches at Jefferson Lab require the accurate theoretical description of the production mechanism in peripheral photoproduction. 
We develop a model for $\pi\Delta$ photoproduction at high energies ($5 \leq \elab \leq 16\units{GeV}$) that incorporates
both the absorbed pion and natural-parity cut contributions.
We fit the available observables, providing a good description
of the energy and angular dependencies of the experimental data.
We also provide predictions for the photon beam asymmetry of charged pions
at $\elab = 9\units{GeV}$ which is expected to be measured by  GlueX and CLAS12
experiments in the near future.
\end{abstract}

\begin{keyword}
\PACS 11.55.Jy \sep 12.39.Jh\\ 
JLAB-THY-17-2580
\end{keyword}
\end{frontmatter}

\section{Introduction\label{sec:introduction}} 
There is mounting evidence for the existence of exotic hadrons that cannot be accommodated within the conventional quark model~\cite{Aaij:2015tga,Alekseev:2009aa,Choi:2003ue,Meyer:2015eta,Esposito:2016noz,Olsen:2017bmm}. Specifically, light flavor hybrid mesons are expected to appear in the spectrum below $\mbox{2 GeV}$~\cite{Dudek:2011bn,Dudek:2013yja}, and to be copiously produced via beam fragmentation in peripheral photoproduction, with photon energies on the order of $10\units{GeV}$~\cite{Close:1994pr,Afanasev:1999rb,Szczepaniak:2001qz}. 
To this end, photoproduction experiments dedicated to the exploration of the hybrid meson spectrum have just begun using the GlueX and CLAS12 detectors at Jefferson Lab (JLab)~\cite{Ghoul:2015ifw}. The success of these experiments relies on the accurate theoretical description of both 
the production mechanism
and the decay of resonances in peripheral photoproduction~\cite{Battaglieri:2014gca}. While resonance decays have been extensively studied in recent years, in view of the forthcoming data, it is necessary to further constrain the production mechanism
~\cite{Yu:2016jfi,Kashevarov:2017vyl,Yu:2011zu,Yu:2017vvp}.
Photoproduction of the light exotic mesons involves the natural-parity ($P(-1)^J =1$) and unnatural-parity ($P(-1)^J=-1$) Regge exchanges that also determine the
photoproduction of pseudoscalar mesons. 
The aforementioned GlueX and CLAS12 experiments have begun a systematic study of pion and $\eta$ production in order to get insight on the production mechanisms~\cite{AlGhoul:2017nbp,Mathieu:2015eia,Nys:2016vjz,Mathieu:2017jjs}. 
The understanding of pion exchange is of particular interest since virtual pions play an important role in various hadronic processes, including the possible formation of hadron molecules~\cite{Guo:2017jvc,Ericson:1993wy}. In peripheral photoproduction, pion exchange dominates forward production.
Because the pion is the lightest meson, it is most sensitive to absorption dynamics, {\it i.e.} final-state interactions~\cite{Irving:1977ea}. 
In this context, we can use the
photon beam asymmetry in charged pion photoproduction
to disentangle the parity
of the exchanged Reggeons and isolate the pion exchange contribution.
In this Letter we predict the beam asymmetry ($\Sigma$) in charged pion photoproduction, associated with production of a $\Delta$ excitation from the proton target. 
The beam asymmetry measurement is free from major systematics,
and is expected to be measured in both the GlueX and CLAS12 
experiments in the near future.
Previous attempts to describe these high-energy observables either fail or do not attempt to reproduce simultaneously the energy and $t$ dependencies~\cite{Yu:2016jfi,Goldstein:1974mr,Clark:1977ce}. 
The aim of this work is to provide a proper account of  these dependencies in the kinematical region relevant to the JLab experiments. Our model is constrained by the differential cross section and beam asymmetry measurements for the reaction $\gamma p\to\pi^+\Delta^{0}$ and $\gamma p\to\pi^-\Delta^{++}$ at 16\units{GeV}~\cite{Boyarski:1967sp,Quinn:1979zp}.

The outline of this Letter is as follows. First, we describe our Regge-theory based model and discuss the necessary absorption corrections. A fit is carried out to the available data. 
The results are extrapolated to lower energies and compared to the available cross section data.
Finally, we provide predictions for the beam asymmetry at JLab energies $\elab = 9\units{GeV}$.

\section{Model\label{sec:model}}
We consider photon beam energies of the order of $\elab = 10\units{GeV}$ which corresponds to 4.4\units{GeV} for the center of mass energy. At low momentum transfer, $\pi\Delta$ photoproduction is dominated by pion exchange at these energies. For $-t \simeq 0.5\units{GeV}^2$ the dynamics are expected to be dominated by natural vector ($\rho$) and natural tensor ($a_2$) exchanges~\cite{Irving:1977ea}. There is also a contribution from the unnatural $b$ exchange that has not been well determined so far. 
We consider a scattering reaction $1+2 \to 3 + 4$ where the particles $1,2,3,4$ denote $\gamma,N,\pi,\Delta$ respectively. The standard Mandelstam variables are $s = (p_1+p_2)^2$ and $t = (p_1-p_3)^2$.
In the Regge pole approximation the asymptotic expression ($s\to \infty$) of the  $s$-channel helicity amplitude for a Regge pole exchange $R$ is given by~\cite{Irving:1977ea,Collins:1971ff,Fox:1973by}
\begin{align}\label{eq:asymptotic}
A^R_{\mu_4\mu_3,\mu_2 \mu_1} &\simeq   \beta^R_{\mu_1 \mu_2 \mu_3 \mu_4}(t) \propagator{R}(s,t) \; .
\end{align}
Here, $\mu_i$ are the $s$-channel helicities, and  $\propagator{R}(s,t)$ is the Regge propagator 
\begin{align}\label{eq:reggeprop}
\propagator{R} = \frac{\pi \alpha_1^R}{2}\frac{\tau_R+e^{-i \pi \alpha^R(t)}}{\sin \pi \alpha^R(t)} \left(\frac{s}{s_0}\right)^{\alpha^R(t)} \; ,
\end{align}
with $\tau_R$ and $\alpha^R_1$ being the signature and slope of the linear Regge trajectory $\alpha^R(t)=\alpha^R_0+ \alpha^R_1 t$, and $s_0 = 1\units{GeV}^2$ a scale factor. 
From unitarity it follows that the residues $\beta^R_{\mu_1\mu_2\mu_3 \mu_4}(t)$ are factorizable, {\it i.e.}  $\beta^R_{\mu_1\mu_2\mu_3\mu_4}(t) =\beta^{R,13}_{\mu_1 \mu_3}(t)\beta^{R,24}_{\mu_2 \mu_4} (t)$. In other words, 
we can factorize the residue in a part originating from the $R13$ vertex and a part from the $R24$ vertex. Angular-momentum and parity conservation determine the non-analytical dependence on $t$. We explicitly define  
$\beta^{R,ij}_{\mu_i \mu_j} (t) = \sqrt{-t}^{\abs{\mu_i - \mu_j}} \betahat^{R,ij}_{\mu_i \mu_j} (t)$ where the reduced residues, $\betahat^{R,ij}_{\mu_i \mu_j}(t)$ are regular in $t$~\cite{Cohen-Tannoudji:1968eoa}. 
In the case at hand, $\beta^R_{\mu_\gamma \mu_N \mu_\Delta}(t) =\beta^{R,\gamma \pi}_{\mu_\gamma}(t)\beta^{R,N\Delta}_{\mu_N \mu_\Delta} (t)$ with $\beta^{R,\gamma\pi}_{\mu_\gamma}(t)\propto \sqrt{-t}$.  That is, in the Regge pole approximation the helicity amplitudes for pseudoscalar meson production vanish near $t=0$.
From overall angular momentum conservation it follows, however, 
that  the $s$-channel helicity amplitude is proportional to the half-angle factor
$\xi_{\mu \mu'}(s,t) = (s(1-z_s)/2)^{\abs{\mu - \mu'}/2}((1+z_s)/2)^{\abs{\mu + \mu'}/2}$,
where $\mu = \mu_1 - \mu_2$ and $\mu' = \mu_3- \mu_4$ is the net helicity flip in the initial and final state, respectively.  
The variable $z_s$ denotes the cosine of the scattering angle in the $s$-channel center-of-mass frame. In the high-energy limit, $z_s \to 1+t'/(2s)$ where $t' = t-t_{z_s=+1}$.
The half-angle factor incorporates the kinematic singularity in $t$, and it asymptotically reduces\footnote{The factor of $s^{\abs{\mu-\mu'}/2}$ ensures that the half-angle factor introduces no additional asymptotic $s$ dependence into Eq.~\eqref{eq:reggeamplitude}.} to $\xi_{\mu\mu'}\xrightarrow[]{s\to\infty}\sqrt{-t}^{\abs{\mu-\mu'}}$.
Matching with the Regge pole form in the asymptotic amplitude given in Eq.~\eqref{eq:asymptotic}, one finds~\cite{Collins:1971ff}
\begin{align}\label{eq:reggeamplitude}
A^R_{\mu_4\mu_3,\mu_2 \mu_1} &= \xi_{\mu \mu'}(s,t) \left(\sqrt{-t}\right)^{-\abs{\mu - \mu'}}\left[ \beta^R_{\mu_1 \mu_2 \mu_3 \mu_4}(t) \propagator{R}(s,t)\right]
\end{align}
The residual analytical dependence in Eq.~\eqref{eq:reggeamplitude} on $t$ coming from the $\betahat$ factors is not predicted by Regge theory. In the following, we use the single-particle exchange model and the data as a guidance to constrain this dependence. Specifically, for the lightest meson on the trajectory $R$, labeled by $e$, the reduced residues are denoted as $\betahat_{\mu_i \mu_j}^{e, i j}(t)$. One expects $\betahat_{\mu_i \mu_j}^{R, i j}(t) \approx \betahat_{\mu_i \mu_j}^{e, i j}(t)$ for small momentum transfer $t$, since the Regge and particle exchange residues coincide at the pole $t \to m_e^2$. 
The residues $\betahat_{\mu_i \mu_j}^{e, i j}(t)$ are proportional to coupling constants $g_{e i j}$ in an effective Lagrangian (see Table~\ref{tab:s_channel_residues}), and in the $s\to \infty$ limit the single-meson exchange amplitude adopts the form
\begin{align}\label{eq:single_particle_amplitude}
A^e_{\mu_\Delta,\mu_N \mu_\gamma} &=   \sqrt{-t}^{\abs{\mu_\gamma }} \sqrt{-t}^{\abs{\mu_N - \mu_\Delta}}\betahat^{e, N \Delta}_{\mu_N \mu_\Delta}(t) \betahat^{e, \gamma \pi}_{\mu_\gamma}(t) \propagator{e}(s,t)\,,
\end{align}
where $\propagator{e} = (s/s_0)^{J_e}/(m_e^2 - t)$ is the propagator of the exchanged particle. 
The Regge propagator in Eq.~\eqref{eq:reggeprop} is normalized such that $\propagator{R} \to \propagator{e}$ for $t\to m_e^2$.
By comparing with Eq.~\eqref{eq:reggeamplitude} one determines the relation between the reduced Regge residues and the elementary couplings, which is summarized in Table~\ref{tab:s_channel_residues}. 
Besides pion exchanges, in the proposed model we include the $\rho$, $a_2$ and $b$ exchanges with signatures $\tau_{\pi,a_2} = +1$ and $\tau_{\rho,b} = -1$. 
The coupling constants are extracted from the corresponding decay widths and are shown in Table~\ref{tab:couplings_from_decays}. We use degenerate $\rho$ and $a_2$ trajectories $\alpha^N \equiv \alpha^{R=\rho,a_2}(t) = 0.9(t-m_\rho^2) + 1$, while for the unnatural $\pi$ and $b$ exchanges we use 
$\alpha^U\equiv\alpha^{R=\pi,b}(t) = 0.7(t-m_\pi^2)$. 
Finally we note that two $\pi \Delta$ channels are related by isospin (neglecting isospin 2),
\begin{align}\label{eq:isospin}
A(\gamma p \to \pi^+ \Delta^0) &=  (A^+ + A^-)/\sqrt{3} \\
A(\gamma p \to \pi^- \Delta^{++}) &= A^+ - A^-
\end{align}
where the $A^G$ ($G$ is the $t$-channel $G$-parity) receive contributions from $\rho$ and $b$, \textit{i.e.}  $A^+ = A^\rho + A^b$ and $a_2$ and $\pi$,  $A^- = A^{a_2} + A^{\pi}$, respectively.

\begin{table}[tbh]
\centering
\caption{The $s$-channel residues from single-meson exchange terms (up to isospin Clebsches-Gordon coefficients). These are obtained by using the Lagrangians in Refs.~\cite{Yu:2011zu,Bellucci:1994eb,Nam:2011np,Yu:2016jfi,Sato:1996gk,Jackson:1964}. All residues must be multiplied by a factor $\sqrt{s_0}^{J_e}$ where $J_e$ is the spin of the corresponding exchange $e$.\label{tab:s_channel_residues}}
\begin{tabular}{|C|L|}
\hline
\betahat^{e, i f}_{\mu_i \mu_f} & \text{Expression} \\
\hline\hline
\betahat^{\pi, \gamma \pi }_{+1}(t) & \sqrt{2} e \\ 
\betahat^{\rho, \gamma \pi}_{+1}(t) & \frac{g_{\rho \pi \gamma} }{2 m_\rho}  \\
\betahat^{b_1, \gamma \pi}_{+1}(t) & \frac{g_{b_1 \pi \gamma}  }{2 m_{b_1}} \\
\betahat^{a_2, \gamma \pi}_{+1}(t) & \frac{g_{a_2 \pi \gamma} }{2 m_{a_2}^2} \\
\hline
\betahat^{\pi, N \Delta}_{+\frac{1}{2} +\frac{3}{2}}(t) & \frac{g_{\pi N\Delta} (m_N + m_\Delta)}{\sqrt{2} m_\Delta} \\
\betahat^{\pi, N \Delta}_{-\frac{1}{2} +\frac{1}{2}}(t) & \frac{g_{\pi   N \Delta} (-m_N^2 + m_N m_\Delta + 2 m_\Delta^2 + t)}{\sqrt{6} m_\Delta^2}\\
\betahat^{\pi, N \Delta}_{+\frac{1}{2} +\frac{1}{2}}(t) & \frac{-g_{\pi   N \Delta} (-m_N^3 - m_N^2 m_\Delta + m_\Delta^3 + 2 m_\Delta t + m_N (m_\Delta^2 + t))}{\sqrt{6} m_\Delta^2} \\
\betahat^{\pi, N \Delta}_{-\frac{1}{2} +\frac{3}{2}}(t) & \frac{-g_{\pi N \Delta}}{\sqrt{2} m_\Delta} \\
\hline
\betahat^{\rho, N \Delta}_{+\frac{1}{2} +\frac{3}{2}}(t) & \frac{-(2 m_\Delta g_{\rho N\Delta}^{(1)} + g_{\rho N\Delta}^{(2)} (m_N - m_\Delta))}{2 m_\Delta^2 }  \\
\betahat^{\rho, N \Delta}_{-\frac{1}{2} +\frac{1}{2}}(t) & \frac{-(2 m_N m_\Delta g_{\rho N\Delta}^{(1)} + g_{\rho N\Delta}^{(2)} (-m_N m_\Delta + m_\Delta^2 + 2t) +2t g_{\rho N\Delta}^{(3)})}{2 \sqrt{3} m_\Delta^3 } \\
\betahat^{\rho, N \Delta}_{+\frac{1}{2} +\frac{1}{2}}(t) & \frac{-(2 m_\Delta g_{\rho N\Delta}^{(1)} + g_{\rho N\Delta}^{(2)} (2m_N - 3 m_\Delta) + 2 g_{\rho N\Delta}^{(3)} (m_N - m_\Delta))}{2\sqrt{3} m_\Delta^3} (-t)\\
\betahat^{\rho, N \Delta}_{-\frac{1}{2} +\frac{3}{2}}(t) & \frac{g_{\rho N\Delta}^{(2)}}{2 m_\Delta^2}\\
\hline
\end{tabular}
\end{table}

\begin{table*}[htb]
\centering
\caption{Decay widths~\cite{reviewofparticlephysics} and respective couplings. Normalizations of the couplings are consistent with Table~\ref{tab:s_channel_residues}.\label{tab:couplings_from_decays}}
\begin{tabular}{|C|C|C|}
\hline
\text{Expression } \Gamma(g) & \Gamma & g \\
\hline\hline
\Gamma_{\rho^\pm \to \pi^\pm \gamma} = g_{\rho\pi\gamma}^2 p^3/(12 \pi m_\rho^2) & 68 \units{keV} & g_{\rho \pi \gamma} = 0.17 \\
\Gamma_{b_1^\pm \to \pi^\pm \gamma} = g_{b_1\pi\gamma}^2 p^3/(12 \pi m_{b_1}^2) & 230 \units{keV} & g_{b_1 \pi \gamma} = 0.24 \\
\Gamma_{a_2^\pm \to \pi^\pm \gamma} = g_{a_2\pi\gamma}^2 p^5/(20 \pi m_{a_2}^4) & 311 \units{keV} & g_{a_2 \pi \gamma} = 0.71 \\
\Gamma_{\Delta \to \pi N} = g_{\pi N \Delta}^2 p^3 (m_N + \sqrt{p^2 + m_N^2})/(12 \pi m_{\Delta}^3) & 116 \units{MeV} & g_{\pi  N \Delta} = 19.16 \\
\hline
\end{tabular}
\end{table*}

The pion exchange is known to be strongly affected by absorption~\cite{Irving:1977ea}, which can be effectively accounted for by a modification of the Regge pole amplitude, known as the ``Williams model", \textit{a.k.a.} ``Poor Man's Absorption'' (PMA)~\cite{Williams:1970rg}.
In  PMA, the $\sqrt{-t}$ factors in the residues that are required by factorization, but not by angular-momentum conservation, are evaluated at the pion pole. Although different in the underlying physics assumptions, the PMA  is equivalent to  a model that adds additional Born terms to the $t$-channel pion exchange~\cite{Fayyazuddin:1970bf,Guidal:1997hy}. 
 We analyze the $\gamma p \to \pi^- \Delta^{++}$ and  $\gamma p \to \pi^+ \Delta^{0}$ differential cross sections and photon-beam asymmetries. In terms of the helicity amplitudes these are given by
 \begin{align}
\dsigdt &= \frac{K}{4} \sum_{\mu_\Delta,\mu_N,\mu_\gamma} \abs{A_{\mu_\Delta,\mu_N \mu_\gamma}}^2, \\
\Sigma \dsigdt &= \frac{K}{4} \sum_{\mu_\Delta,\mu_N} 2 \Re{A_{\mu_\Delta,\mu_N \mu_\gamma=+1} A_{\mu_\Delta,\mu_N \mu_\gamma=-1}^*}, \\
\dsigdtperppara &= \frac{K}{4} \sum_{\mu_\Delta,\mu_N} \abs{A_{\mu_\Delta,\mu_N \mu_\gamma=+1} \pm A_{\mu_\Delta,\mu_N \mu_\gamma=-1}}^2,
\end{align}
with  $K = (64 \pi s p_{s12}^2)^{-1}$ and $\txdsigdtperppara$ the differential cross section for photon polarizations perpendicular/parallel to the reaction plane.
The unpolarized differential cross section is denoted by $\txdsigdt$ and $\Sigma$ is the photon beam asymmetry.

\begin{figure}[tbh]
\centering
\includegraphics[height=6cm,keepaspectratio]{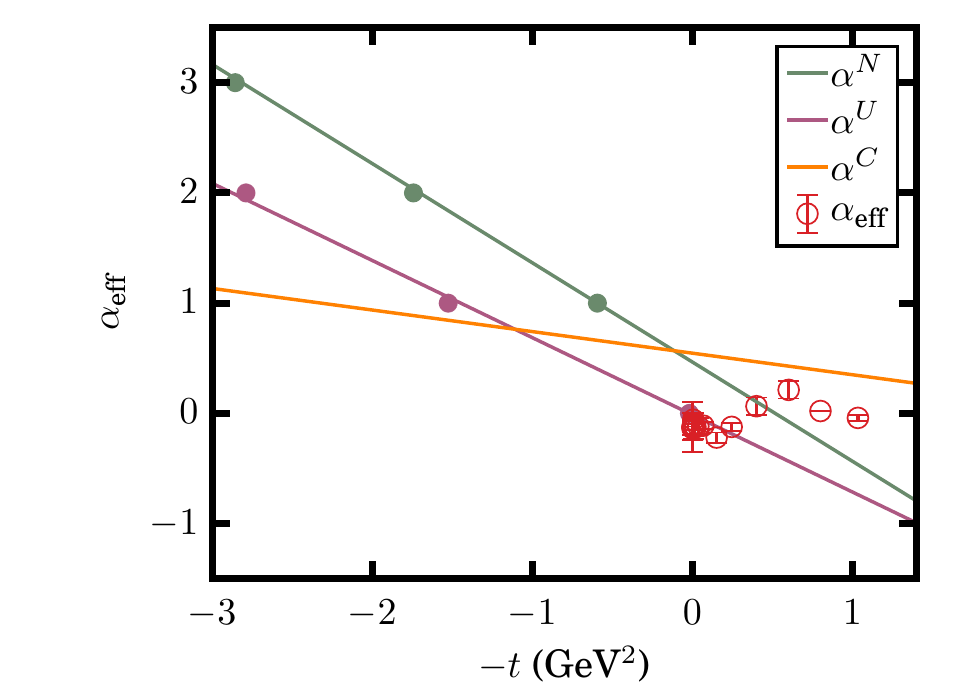}
\caption{The $t$ dependence of some selected trajectories. The effective trajectory of the cross section $\alpha_\text{eff}(t)$ (red) is extracted with the aid of a fit to the data with the function in Eq.~\ref{eq:fit.function}.
The green and purple solid curves illustrate the Regge trajectories used in this work (see text), together with observed particles. The orange line depicts the $\rho \otimes \pomeron$ or $a_2 \otimes \pomeron$ cut trajectory.\label{fig:alphaeff}}
\end{figure}

We first extract the effective trajectory $\alpha_\text{eff}(t)$ by studying $s$ dependence at fixed $t$ of the available unpolarized cross sections for $\pi^- \Delta^{++}$ photoproduction.
We use an asymptotic approximation for the $s$ dependence
\begin{align} \label{eq:fit.function}
\dsigdt \simeq f(t) s^{2\alpha_\text{eff}(t) - 2}\,.
\end{align}
The results for the fitted $\alpha_\text{eff}$ are shown in Fig.~\ref{fig:alphaeff} as a function of $t$.
As expected, we find that  
pion exchange ($\alpha_\text{eff} \simeq 0$) dominates at small $-t$, while  
 natural exchange contributions become important at $-t \geq 0.5 \units{GeV}^2$ resulting in  $\alpha_\text{eff}(t \geq 0.5 \units{GeV}^2 ) \simeq 0.5$. Overall, however, $\alpha_\text{eff}(t)$ is not as steep as compared to the expectation from a pure Regge pole, indicating the presence Reggeon-Pomeron rescattering or daughter poles, which in general flattens the $t$ dependence. Guided by this observation, we consider two scenarios: ($i$) the $\rho$ and $a_2$ exchanges are described as pure Regge poles, and ($ii$) we include final state interaction corrections. 
In the latter, we replace the pole trajectory by a cut trajectory $\alpha^N(t) \to \alpha^C(t)=\alpha^N_0+\alpha^\pomeron_0 -1 + t(\alpha^N_1 \alpha^\pomeron_1)/(\alpha^N_1 + \alpha^\pomeron_1) $. For the Pomeron we use $\alpha^\pomeron(t)=1.08 + 0.25 t$~\cite{List:2009pb}. In addition, the explicit calculation of the absorption correction gives an additional factor of $\left(\ln s/s_0\right)^{-1}$~\cite{Irving:1977ea}, which we include.
Even though the cut trajectory and effective trajectory do not fully match 
 (see Fig.~\ref{fig:alphaeff}), the remaining factors in the Regge amplitude ({\it i.e} the half-angle factor and the extra $\ln s/s_0$ dependence) ultimately results in a good agreement with the data (see Fig.~\ref{fig:observables}).
While in the Reggeon-Pomeron cut model for the $\rho$ and $a_2$, the connection between the Regge and single-particle residues is lost, we still use the same parametrization since it provides enough freedom in the fit. We verified that alternative parameterizations for the $t$ dependence of the residues of the natural exchanges do not change the conclusions of the following analysis, nor do they significantly alter the predictions for JLab energies.

The Regge propagator in Eq.~\eqref{eq:reggeprop} contains ghost poles which must be canceled by zeros in the residues. Exchange degeneracy (EXD) forces these zeros to appear in the residue of the EXD partner as well, implying zeros in the amplitude. The latter are referred to as nonsense wrong-signature zeros (NWSZ).
Since EXD does not in general hold for the overall residue in photoproduction reactions, for each individual Reggeon we only remove those ghost poles that are closest to the physical region under consideration, without including NWSZ. 
In particular, we remove the ghost poles\footnote{In removing these ghost poles we respect the normalization of the residues on the lightest mass pole of the EXD trajectories.} at spins $\alpha=-2$ for $\pi$, $\alpha=-1$ for $b$, $\alpha=-1$ for $\rho$ and $\alpha=0,-2$ for the $a_2$. 
At this point, it is worth mentioning that NWSZ are not favored by the data.
Absence of such zeros was noted in the analysis of Yu~\textit{et al.}~\cite{Yu:2016jfi}, where to fill in the dips, the authors replace the signature factors of the $\rho$ and $a_2$ with a different phase. While the physics behind such a phase is not well justified in principle\footnote{EXD is an equality between two Reggeons. The constant and rotating phases are in principle obtained when two Regge contributions with equal residues are added or subtracted.}, the effect of this substitution is to remove the NWSZ in both contributions.

The unnatural and natural contributions have an overall exponential factor which accounts for the phenomenological falloff  at large values of $-t$. Explicitly,
\begin{subequations}
\begin{align}
\betahat^{R=\pi}_{\mu_\gamma \mu_N \mu_\Delta}(t) =& \:c_{\pi} \betahat^{e=\pi}_{\mu_\gamma \mu_N \mu_\Delta}(t) e^{b_{U}t} (\alpha(t)+2)/2\,, \\
\betahat^{R=b}_{\mu_\gamma \mu_N \mu_\Delta}(t) =&\:c_{\pi}\betahat^{e=b}_{\mu_\gamma \mu_N \mu_\Delta}(t) e^{b_{U}t} (\alpha(t)+1)\,, \\
\betahat^{R=\rho}_{\mu_\gamma \mu_N \mu_\Delta}(t) =&\: \betahat^{e=\rho}_{\mu_\gamma \mu_N \mu_\Delta}(t)e^{b_{N}t} (\alpha(t)+1)/2\,, \\
\betahat^{R=a_2}_{\mu_\gamma \mu_N \mu_\Delta}(t) =&\: \betahat^{e=a_2}_{\mu_\gamma \mu_N \mu_\Delta}(t)e^{b_{N}t} \alpha(t) (\alpha(t)+2)/3 \,.
\end{align}
\end{subequations}
The $\betahat$ on the left and right hand side of the above equations are the Regge and single-particle residues, respectively. We introduced an additional factor $c_{\pi}$ in order to allow small deviations from the estimated pion couplings.
We require $\betahat^{a_2, p \Delta^{++}}(t) = \sqrt{s_0}\betahat^{\rho, p \Delta^{++}}(t)$ and $\betahat^{b, p \Delta^{++}}(t) = \sqrt{s_0}\betahat^{\pi, p \Delta^{++}}(t)$ up to the ghost killing factors.
For the photon vertex we use the radiative decay couplings from Table~\ref{tab:couplings_from_decays}. 

\section{Results\label{sec:results}}
For the two isospin channels $\pi^+ \Delta^0$ and $\pi^- \Delta^{++}$ data are available for the differential cross sections, the polarization cross sections and the beam asymmetries at a single energy $\elab = 16\units{GeV}$. High-energy data within $5 \leq \elab < 16\units{GeV}$ are available for $\pi^- \Delta^{++}$ only~\cite{Boyarski:1967sp,Quinn:1979zp}. For definite parity exchanges, the polarization cross sections are useful, since they are sensitive to a given naturality in the $t$-channel. Specifically, $\dsig_\perp$ ($\dsig_\parallel$) are determined by natural (unnatural) contributions~\cite{Mathieu:2015eia}, respectively. Thus, knowledge of $\dsig_\parallel$ allows us to study $\pi$ exchange in isolation. It should be noted, however, that absorption effectively changes  the naturality of the $\pi$ exchange and PMA specifically results, in the forward region, in an equal contribution to both naturalities. Hence, $\dsig_\perp$ also contains contributions from absorbed pion exchanges.

\begin{table}[tbh]
\centering
\caption{Fitted parameters for the two models. In the pole model, all exchanges are pure Regge poles. In the cut model, $\rho$ and $a_2$ contributions are Reggeon-Pomeron cuts.\label{tab:fit_params}}
\begin{tabular}{|c|c|c|}
\hline
& Pole model & Cut model \\
\hline\hline
$c_\pi$ & $\phantom{-}1.06$ & $\phantom{-}1.04$ \\
$b_{U} (\text{GeV}^{-2})$ & $\phantom{-}0.06$ & $\phantom{-}0.14$\\
$b_{N} (\text{GeV}^{-2})$ & $-0.42$ & $-2.12$\\
$g^{(1)}_{\rho N \Delta}$ & $-48.2$ & $-370.8$ \\
$g^{(2)}_{\rho N \Delta}$ & $-52.4$ & $-242.4$ \\
$g^{(3)}_{\rho N \Delta}$ & $\phantom{-}40.2$ & $-139.0$ \\
 \hline
\end{tabular}
\end{table}

From the analysis of radiative decays and Table~\ref{tab:s_channel_residues}, we find
$\beta^{a_2, \gamma \pi}_{+1}/\beta^{\rho, \gamma \pi}_{+1} = 1.82$ and
$\beta^{\pi, \gamma \pi}_{+1}/\beta^{b, \gamma \pi}_{+1} = 4.38$.
Hence, the $\rho$ and $b$ contributions are suppressed with respect to their opposite signature partners. In Refs.~\cite{Goldstein:1974mr,Yu:2016jfi}, the authors used a value of $3$ for both ratios.
The obtained $c_\pi$ value is consistent with unity and is mainly fixed by the $\dsig_\parallel$ data, which is dominated by $\pi$ exchange. 
 Observing a significant difference in $\dsig_\perp$ between the two isospin channels in the region around $\sqrt{-t} = 0.4\units{GeV}$, one concludes that the $\rho$ and $a_2$ contributions must have a rather strong $t$ dependence. Indeed, one can exclude the presence of strong variations in $t$ in the pion residue due to the rather featureless $t$ dependence of $\dsig_\parallel$. Since the $\rho N \Delta$ couplings are not well constrained, we obtain them from a fit. 
The PMA model reproduces well the forward behavior, thereby correctly matching the natural and unnatural contributions. Indeed, all natural contributions stemming from $\rho$ and $a_2$ exchanges are suppressed in the forward direction by the $\sqrt{-t}$ factors. 
By neglecting the $b$ exchange contribution, the difference between the isospin channels is attributed to the interference of the $\rho$ with the $a_2$ and $\pi$ terms. If the $\rho$ exchange has a NWSZ at $t = -0.55\units{GeV}^2$, 
 $A^+\approx 0$ and the two isospin channels would coincide in this region. This is not observed in the data. Hence, the residues of $\rho$ cannot contain NWSZ within the pure Regge pole model. The NWSZ in the $\pi^+ p \to \pi^0 \Delta^{++}$ cross section must therefore be accounted for by the $\rho \pi \pi$ residue. A similar lack of NWSZ in the $\rho$ exchange in photoproduction reactions was found in Ref.~\cite{Nys:2016vjz}, where a detailed mapping of the $t$ dependence of the residues was carried out through the use of finite-energy sum rules.

\begin{figure*}[tbh]
\centering
\begin{subfigure}{.43\textwidth}
  \centering
  \includegraphics[height=5cm,keepaspectratio]{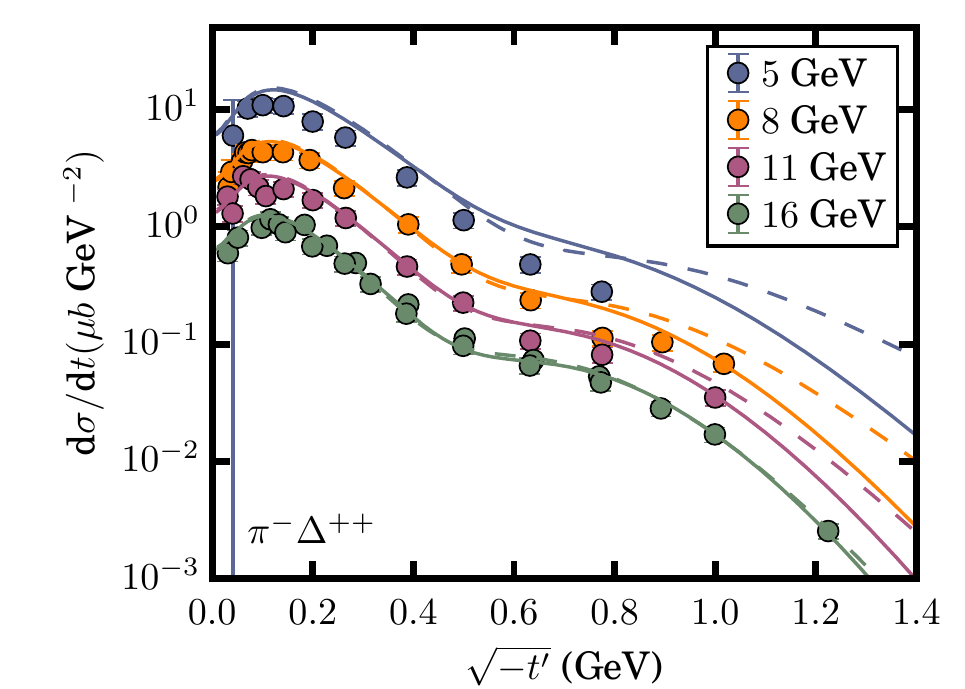}
\end{subfigure}%
\begin{subfigure}{.3\textwidth}
  \centering
  \includegraphics[height=5cm,keepaspectratio]{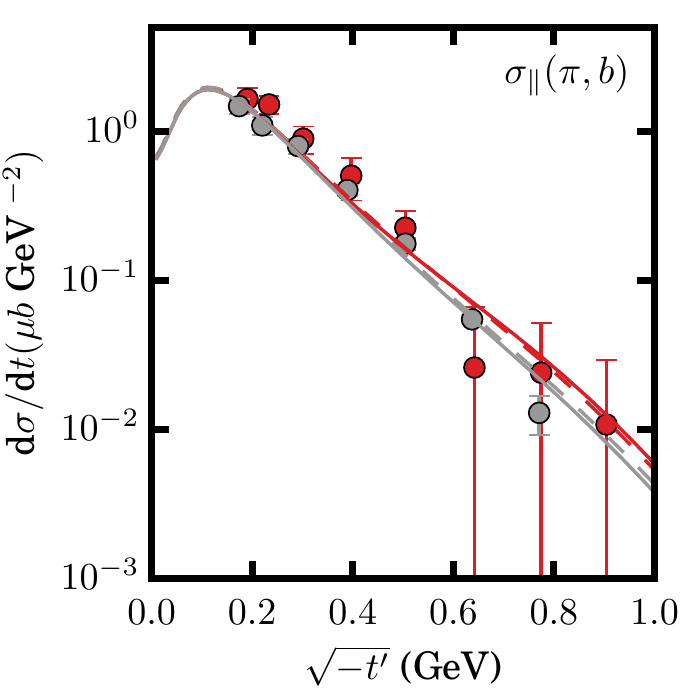}
\end{subfigure}%
\begin{subfigure}{.3\textwidth}
  \centering
  \includegraphics[height=5cm,keepaspectratio]{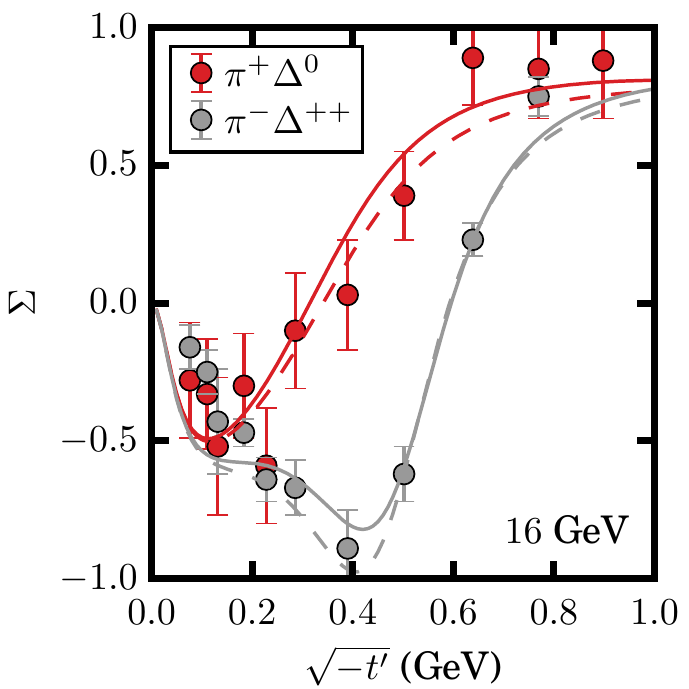}
\end{subfigure}%

\begin{subfigure}{.43\textwidth}
  \centering
  \includegraphics[height=5cm,keepaspectratio]{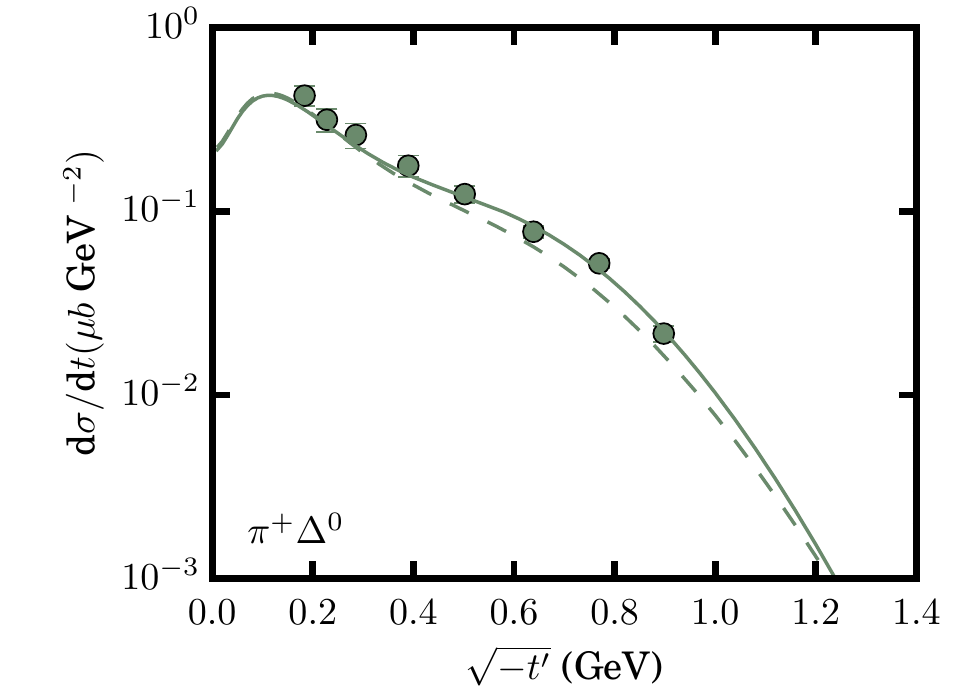}
\end{subfigure}%
\begin{subfigure}{.3\textwidth}
  \centering
  \includegraphics[height=5cm,keepaspectratio]{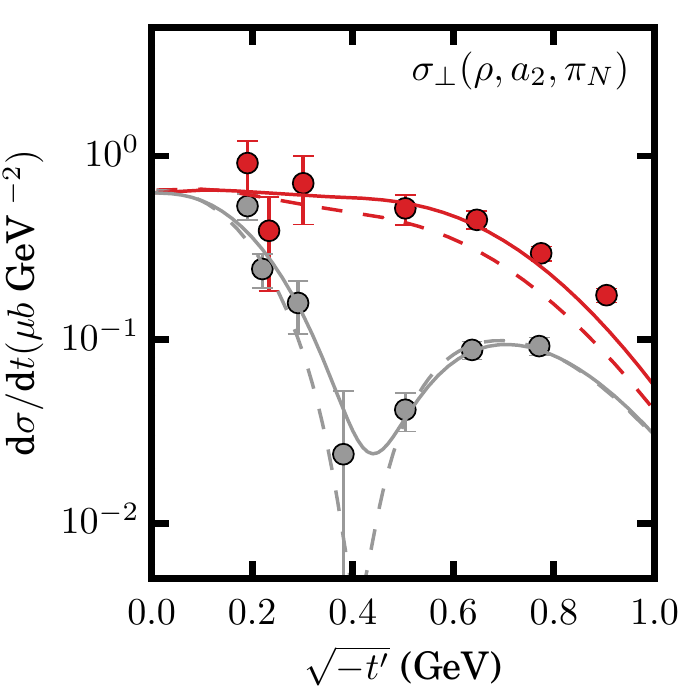}
\end{subfigure}%
\begin{subfigure}{.3\textwidth}
  \centering
  \includegraphics[height=5cm,keepaspectratio]{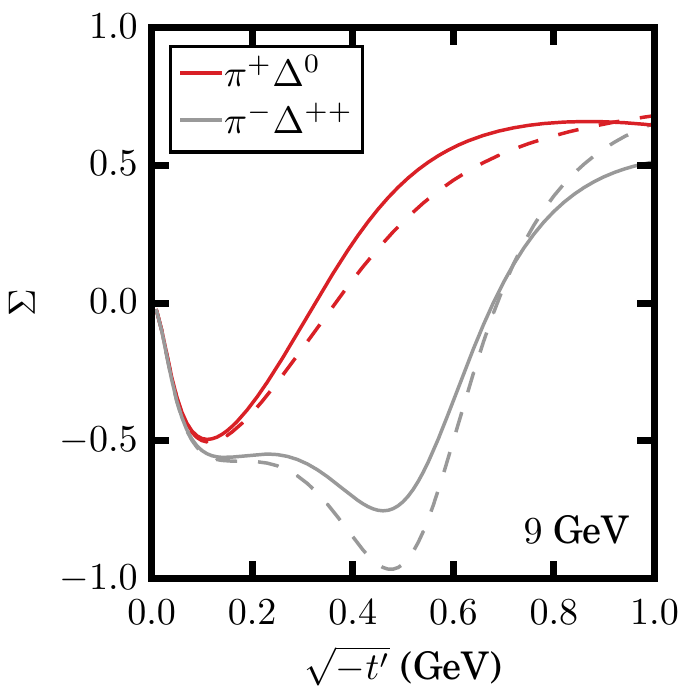}
\end{subfigure}

\caption{The comparison of pole (dashed) and cut (solid) models with the available (unpolarized) differential cross section and beam asymmetry data from Refs.\cite{Quinn:1979zp,Boyarski:1967sp,Boyarski:1968dw}. In the bottom right panel, we show the predictions for $\elab = 9\units{GeV}$ which is relevant for the GlueX experiment. The data and model for $\gamma p \to \pi^-\Delta^{0}$ have been rescaled by a factor of $3$ to compensate the overall isospin coefficient in Eq.~\eqref{eq:isospin}.\label{fig:observables}}
\end{figure*}

The fits are constrained with all of the available $\elab = 16\units{GeV}$ data, leaving the $\elab = 5,8,11 \units{GeV}$ cross section data as a prediction and model validation. The results of the fits are shown in Fig.~\ref{fig:observables}.
The fitted parameters are given in Table~\ref{tab:fit_params}. 
Even though both the pure pole and pole-plus-cut model describe the data rather well, we observe quite a sensitivity in the normalization of the $\rho N \Delta$ couplings. Thus an independent estimate of these parameters would be very important. In our fits this is driven by the large difference in the observed beam asymmetry for the two isospin channels. The model in Ref.~\cite{Yu:2016jfi}  was not given as much freedom in a fit to the data as in the current analysis, but rather the couplings were constrained by symmetry arguments. However, from a comparison of the presented model with the one in Ref.~\cite{Yu:2016jfi}, it becomes clear that pure pole-like contributions with natural size couplings are not able to reproduce the aforementioned behavior. The new experiments at JLab will be able to address this complex feature. The main deficiency of the poles-only fit is that it overestimates the $s$ dependence of the $\pi^- \Delta^{++}$ cross section at large $-t$. A natural-parity cut contribution coincides with the observed energy dependence, except for the $\elab = 5$ GeV data. At such low energies, daughter and additional cut contributions are expected.

We can now predict the beam asymmetry at JLab energies of $\elab = 9\units{GeV}$ as shown in Fig.~\ref{fig:observables}. The predicted observable appears rather similar to the SLAC data at $\elab = 16\units{GeV}$~\cite{Boyarski:1968dw}. The underlying dynamics can be interpreted in the following way. At high $-t$, $\Sigma\approx +1$ indicates dominance of natural exchanges. As $-t$ becomes smaller, pion exchanges dominate the forward region, which is reflected by $\Sigma \to -1$. For $t'\to 0$, one expects $\Sigma = -1$ for purely factorizable exchanges, since the pion remains the dominant contribution up to extremely forward angles. However, the effect of $\Sigma \to 0$ indicates the presence of additional non-pole terms of equal parity in the $t$-channel, as successfully included by the PMA model.

\section*{Acknowledgements}
This material is based upon work supported in part by the Research Foundation Flanders (FWO-Flanders), the U.S.~Department of Energy, Office of Science, 
Office of Nuclear Physics under contract DE-AC05-06OR23177, DE-FG0287ER40365.
National Science Foundation under Grants PHY-1415459 and PHY-1205019, the IU Collaborative Research Grant,
PAPIIT-DGAPA (UNAM, Mexico) grant No.~IA101717,
CONACYT (Mexico) grant No.~251817, 
and Red Tem\'atica CONACYT de 
F\'{\i}sica de Altas Energ\'{\i}as (Red FAE, Mexico).

\newpage
\bibliographystyle{elsarticle-modified}
\bibliography{biblio}

\end{document}